# Experimental multi-scale characterization of mode-II interlaminar fracture in geometrically scaled stitched and unstitched resin-infused composites

Dawson Ozborn[1], Jackob Black[2], Wayne Huberty[3], Christopher Bounds[4], Han-Gyu Kim[5]

*Mississippi State University, Mississippi State, MS 39762, USA*

**This work is focused on investigating the impact of out-of-plane stitches on enhancing mode-II interlaminar fracture toughness (or energy) and characterizing damage progression and crack arrestment in stitched resin-infused composites. For the experimental work, End-Notched Flexure (ENF) quasi-isotropic specimens were manufactured using ±45 non-crimp carbon-fiber fabrics through a resin-infusion process. Both stitched and unstitched specimen sets were designed for comparison. For a size effect study, the ENF specimens were geometrically scaled with three scaling levels. Based on the load-displacement data (i.e., global analysis), the fracture energy of the specimen material was analyzed using the compliance calibration method and a size effect theory. The fracture energy values were compared between the stitched and unstitched cases to characterize the enhanced fracture toughness of stitched composites. For local analysis, two types of digital image correlation (DIC) systems were employed: microscopic and macroscopic (i.e., coupon-scale) DIC systems. By analyzing in-plane displacement through the thickness, separation development was characterized along predicted fracture process zones. The impact of out-of-plane stitches on separation propagation along fracture process zones was discussed based on the DIC analysis. This work will contribute to developing a high-fidelity damage model for stitched resin-infused composites in the form of a traction-separation for high-speed aircraft applications.**

## I. Introduction

Modern aircraft designs have extensively adopted composites for primary structures to achieve lightweight. The Boeing 787 (a subsonic commercial airliner) and the Lockheed F-35 (a supersonic fighter) are representative examples [1, 2]. Composite fuselage and wing structures in high-speed aircraft, however, are subjected to the extreme environments induced by aerothermodynamic couplings, which could lead to stress amplification and mode-II interlaminar failure [3]. To address this issue, the work herein is focused on investigating the impact of out-of-plane stitches on enhancing mode-II interlaminar fracture toughness (or energy) [4]. Characterizing damage progression and crack arrestment in stitched resin-infused composites on the microscopic scale using a through-thickness deformation analysis method [5] is of primary interest.

For experimental characterization of the mode-II interlaminar fracture energy of composite materials, Davidson et al. [6, 7] proposed the compliance calibration method (CCM) based on the Linear Elastic Fracture Method (LEFM). This method has been adopted in the ASTM D7905/D7905M-19el testing manual [8]. Salviato et al. [9], however, experimentally characterized mode-II quasi-brittle fracture processes in geometrically scaled composites and showed that this LEFM-based method could estimate multiple fracture energies for a single material (i.e., higher fracture energies with higher scaling factors). To address this issue, they applied Bažant's type-II size effect law [10] to the experimental load-displacement data in conjunction with numerical J-integral analysis [11] and obtained the fracture energy of the specimen material as a single value. In this paper, the fracture energy values were evaluated based on both CCM and the size effect law and were compared between the analysis methods and also between the stitched and unstitched cases.

The work herein will be extended to developing a high-fidelity damage model to predict the structural life of stitched resin-infused composites for high-speed aircraft under aerothermodynamic coupling [3,12]. The damage model will be built based on traction-separation laws and cohesive zone modeling approaches. Some of the local separation analysis results are presented for both the stitched and unstitched cases at the end of this paper.

## II. Specimen Design and Manufacturing

### A. Specimen details

The specimens were laid up using Saertex Class-75 non-crimp fabrics which consist of [0/90] carbon-fiber layers. Based on the ASTM specifications for End-Notched Flexure (ENF) specimens [8], the geometric dimensions were designed. It needs to be noted that given the characteristics of the material type, quasi-isotropic layups were adopted in this work instead of unidirectional layups. For a size effect study, the ENF specimens were geometrically scaled in three levels as shown in Fig. 1. The dimensions of the Size-3 specimens were determined based on the ASTM specifications while the thickness value was larger than the ASTM specification to design symmetric layups in the smallest set (i.e., the Size-1 set). Having the Size-3 dimensions as a baseline, the thickness, gauge length, and initial crack length values were scaled down for the Size-1 and Size-2 specimens. The target scaling factors were 1/2 and 1/4 for the Size-2 and the Size-1 set, respectively. Scaling was not made for the width and the nominal width of all the size sets was 25 mm. To induce precracks in the specimens, DuPont Teflon® FEP films (12.5-µm thick) were embedded in the midplanes as inserts. The layups for the Size-1, Size-2, and Size-3 sets are [-45/45/90/0]$_S$, [-45/45/90/0]$_{2S}$, and [-45/45/90/0]$_{4S}$, respectively.

### B. Manufacturing details

Six panels (16.5 in x 16.5 in squares) were laid up using the fabrics. Half of the layups were stitched through the thickness, while the others stayed unstitched for comparison. The stitching process is shown for a Size-3 panel in Fig. 2. The layups (see Fig. 2a) were stitched using a Ferdco Juki Pro 2000h with Vectran 1200 denier filament yarns (see Fig. 2b). The stitches were made along the predicted fracture process zones (FPZs) between the initial crack tips and loading points. The Size-1, Size-2, and Size-3 sets had two, four, and eight stitch seams, respectively, with 2.5 mm spacing. The stitched and unstitched layups were infused with API 1078 resin using the Vacuum Assisted Resin Transfer Molding (VARTM) method as shown in Fig. 3. An APT SRD series automated resin-infusion system (see Fig. 3a), a two-part epoxy-amine resin system developed for structural applications for composites in aircraft, was employed for the VARTM process. The resin-infused panels (see Fig. 3b for a Size-3 panel) were cut using a tile saw to have the specimens in the designed dimensions (see Fig. 1).

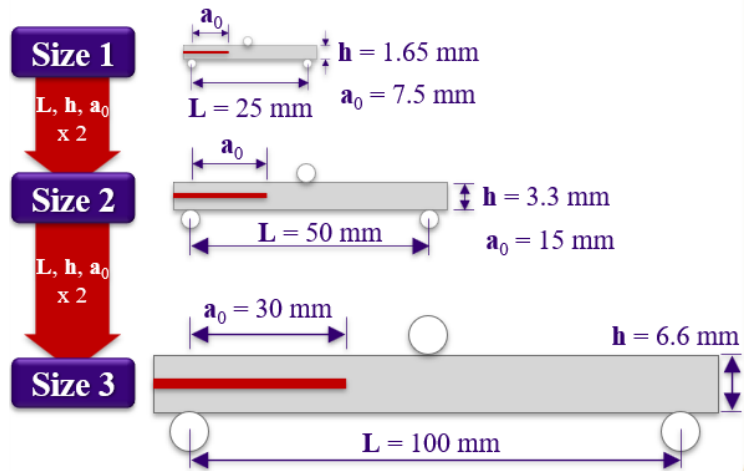

**Figure 1. Nominal dimensions of geometrically scaled specimens**

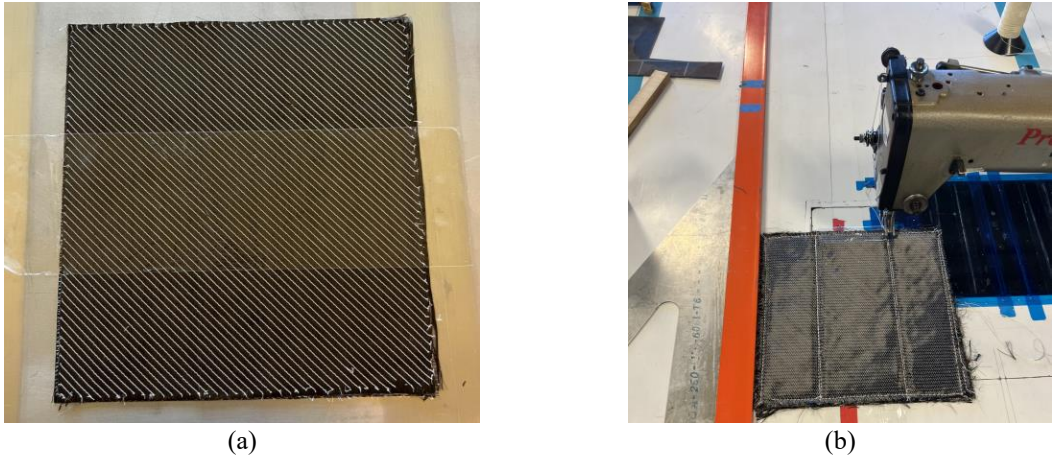

(a)       (b)

**Figure 2. Stitching process. (a) A layup of Saertex Class-75 non-crimp fabrics with a DuPont Teflon® FEP film on the midplane. (b) Stitching on the layup with a Ferdco Juki Pro 2000h.**

## III. Experimental Work

To obtain data on both global and local fracture behaviors, three-point bending tests of the ENF specimens were performed according to the ASTM D7905/D7905M-19el specification for mode-II interlaminar fracture [8]. Following is a detailed account of the experimental setup employed as well as analysis methodology for the collected data.

### A. Experimental setup

For multi-scale characterization of damage progression and arrestment, microscopic and macroscopic (i.e., coupon-scale) Digital Image Correlation (DIC) systems were employed as shown in Fig. 4. For the microscopic system (see Fig. 4a), a Psylotech μTS testing frame and an Olympus BXFRM microscope with a 12-MP machine vision camera were used along with a Correlated Solutions VIC 2D package. The field of view (FOV) of the microscopic tests was 10 mm x 7.3 mm and thus focused on the FPZs between the initial crack tips and loading points. This system was intended to capture crack initiation, progression, and arrestment processes in the microscopic scale. For the Size-3 sets, on the other hand, the FOV was not large enough to cover the whole FPZs. To address this issue, a coupon-scale system was instead applied for the sets as shown in Fig. 4b by replacing the microscope with a 12-MP macroscopic lens, which can cover the entire specimen surfaces.

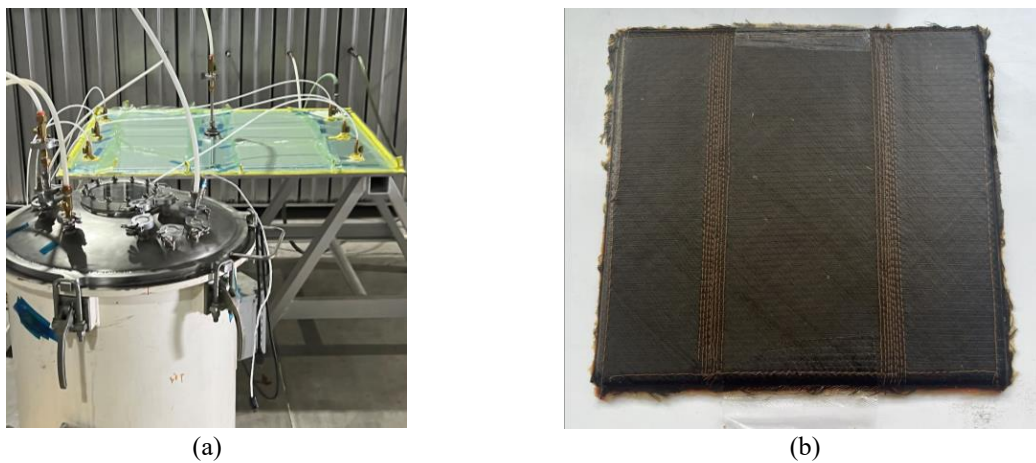

(a)       (b)

**Figure 3. VARTM process for a Size-3 panel. (a) A VARTM setup using an APT SRD series automated resin-infusion system. (b) A stitched Size-3 panel immediately after the resin-infusion process.**

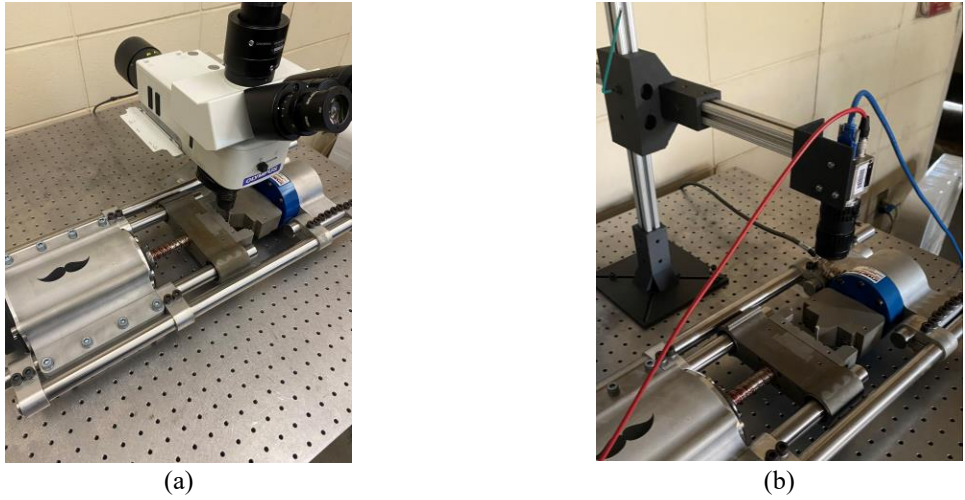

**Figure 4: Experimental setup. (a) Microscopic setup. (b) Coupon-scale setup.**

## B. Analysis methodology

Experimental works were focused on characterizing both global and local fracture behaviors. The global analysis was intended to characterize the mode-II interlaminar fracture energy of the material by analyzing load-displacement curves based on CCM and a size effect theory. The details on CCM and size effect analysis can be found in Refs. [8] and [9], respectively. A CCM analysis of an unstitched Size-3 specimen is illustrated in Fig. 5. The three load-displacement curves of the specimen (see Fig. 5a) were obtained by changing the initial crack lengths to $a_0$, $a_1$, and $a_2$, while major parameters for fracture energy analysis were obtained from the linear regression analysis of the corresponding compliance values. Another fracture energy set was obtained using Bažant's type-II size effect law [13]. For the size effect analysis, the law was applied to the load-displacement data and the J-integral analysis of normalized fracture energy for a nominal Size-2 specimen was made using the Abaqus/Standard solver [14]. It needs to be noted that stitched composites show two different fracture process regimes: quasi-brittle fracture and crack arrestment regimes as shown in Fig. 6. The quasi-brittle fracture regime begins at the beginning of the tests and ends at a sudden load drop (or a snapdown in the curve), where the fracture mechanism changes and the crack arrestment regime begins. In this study, the fracture energies of the stitched and unstitched composites were compared only in the quasi-brittle fracture regime. The local analysis, on the other hand, was made to characterize separation development along the FPZs using the DIC data. The separation values were obtained by analyzing in-plane displacements through the thickness. The details on the through-thickness deformation analysis method can be found in Ref. [5]. The separation analysis for the stitched composites was also made only in the quasi-brittle fracture regime. More studies will be done for the crack arrestment regime in future work.

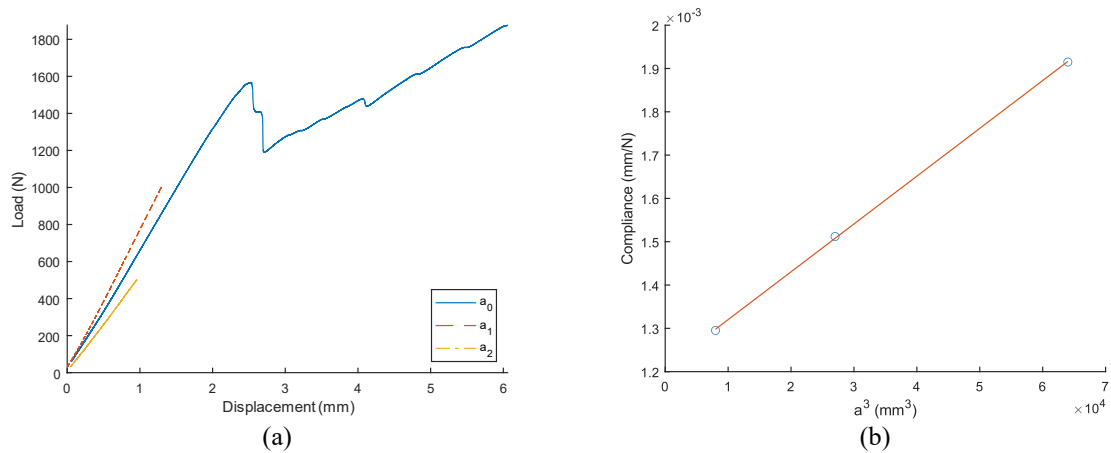

**Figure 5.  CCM analysis of an unstitched Size-3 specimen. (a) Load-displacement curves of the specimen under different crack lengths $a_0$, $a_1$, and $a_2$. (b) Linear regression analysis of the compliance data.**

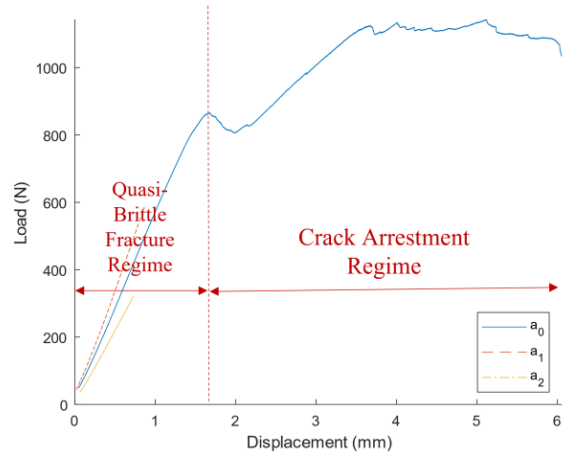

**Figure 6. Load-displacement curve of a stitched Size-2 specimen**

## IV. Experimental Result

In this section, the experimental results are presented and discussed for global and local fracture behaviors. The analysis and discussions for the global behavior are focused on comparing the mode-II interlaminar fracture energy values of the specimen material between the stitched and unstitched cases and also between the CCM and size effect analysis cases. The discussions on the local behavior, on the other hand, focuses on the impact of out-of-plane stitches on separation propagation along the FPZs.

### A. Global fracture behaviors

For global fracture behavior analysis, the load-displacement curves of all the specimens are presented in Fig. 7. It needs to be noted that the tests for the stitched cases (see Fig. 7b) were stopped at certain displacements prior to complete collapse since this paper is focused on the quasi-brittle regime. The unstitched specimens (see Fig. 7a) showed relatively consistent curves within each size set. The Size-2 and Size-3 sets showed clear snapdown behaviors as post-peak response; however, the size-1 set (the smallest one) showed more irregular post-peak behaviors. The load-displacement curves of the stitched specimens (see Fig. 7b), on the contrary, were highly inconsistent between specimens. Also, relatively weaker snapdown behaviors were observed from the curves due to the crack arrestment near the stitches. The normalized peak loads $\sigma_{Nc}$ are tabulated with the fracture energy values $G_{f,CCM}^{(II)}$ obtained from the CCM analysis in Table 1. The normalized peak loads $\sigma_{Nc}$ were obtained by dividing the peak loads of the specimens by the corresponding cross-sectional areas. The values $\sigma_{Nc}$ and $G_{f,CCM}^{(II)}$ in the table are the average values. With size increase, both the unstitched and stitched cases showed an increase in $\sigma_{Nc}$ but a decrease in $G_{f,CCM}^{(II)}$ except for the unstitched Size-1 set. Compared to the unstitched specimens, the stitched specimens showed higher $\sigma_{Nc}$ and $G_{f,CCM}^{(II)}$ for the Size-2 and Size-3 cases but lower values for the Size-1 set. The different between the values $\sigma_{Nc}$ and $G_{f,CCM}^{(II)}$ of the stitched and unstitched specimens at the quasi-brittle regime could imply that stitches on the FPZs make impacts on crack initiation and propagation before the cracks are arrested at the stitches. This phenomenon will be further investigated in future work.

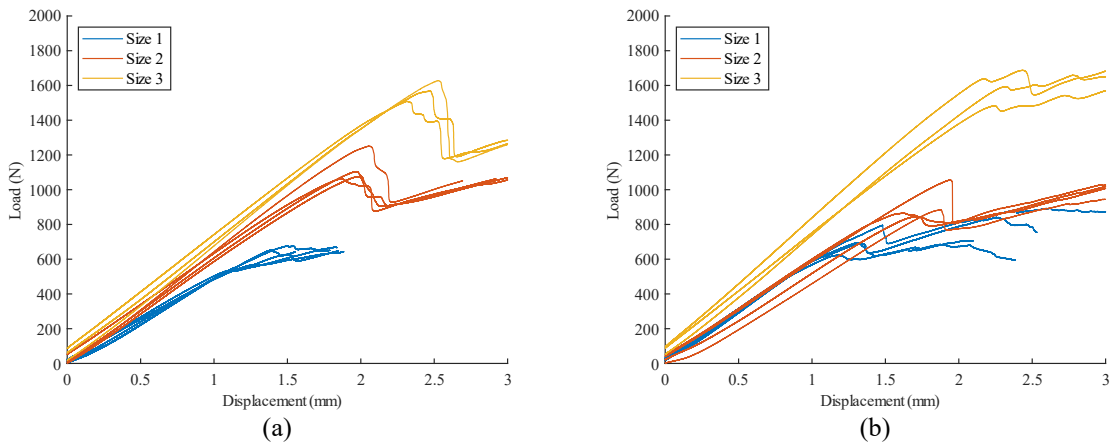

**Figure 7. Load-displacement curves of all the specimens. (a) Unstitched specimens. (b) Stitched specimens.**

**Table 1. CCM analysis results**

| | Unstitched | | | Stitched | |
|---|---|---|---|---|---|
| Size | $\sigma_{Nc}$ (MPa) | $G_{f,CCM}^{(II)}$ (N/mm) | Size | $\sigma_{Nc}$ (MPa) | $G_{f,CCM}^{(II)}$ (N/mm) |
| 1 | 15.15 | 1.59 | 1 | 13.98 | 0.88 |
| 2 | 10.19 | 1.20 | 2 | 13.14 | 1.58 |
| 3 | 9.02 | 1.49 | 3 | 9.30 | 1.85 |

To obtain a single fracture energy value as a material property, Bažant's type-II size effect law was applied to the load-displacement data following the work in Ref. [9]. The analysis results are tabulated in Table 2 and the size effect curves are presented in Fig. 8. As shown in the table, the pseudo-plastic limit $\sigma_0$, transitional characteristic length $h_0$, fracture energy $G_f^{(II)}$ were all similar between the unstitched and stitched cases, while the effective FPZ size $c_f$ was larger for the stitched case. The increased $c_f$ induced by stitching could substantiate the argument that stitches along the FPZs could affect crack initiation and propagation prior to crack arrestment at the stitches. The size effect curves showed that the transition between the pseudo-plastic fracture and LEFM regimes was well captured by the Size-1 and Size-2 specimens for both the unstitched and stitched cases (see Figs. 8a and 8b, respectively), while the Size-3 sets are close to the LEFM limits.

**Table 2. Size effect analysis results**

| Items | $\sigma_0$ (MPa) | $h_0$ (mm) | $c_f$ (mm) | $G_f^{(II)}$ (N/mm) |
|---|---|---|---|---|
| **Unstitched** | 19.95 | 1.92 | 4.53 | 2.64 |
| **Stitched** | 17.90 | 2.32 | 5.46 | 2.56 |

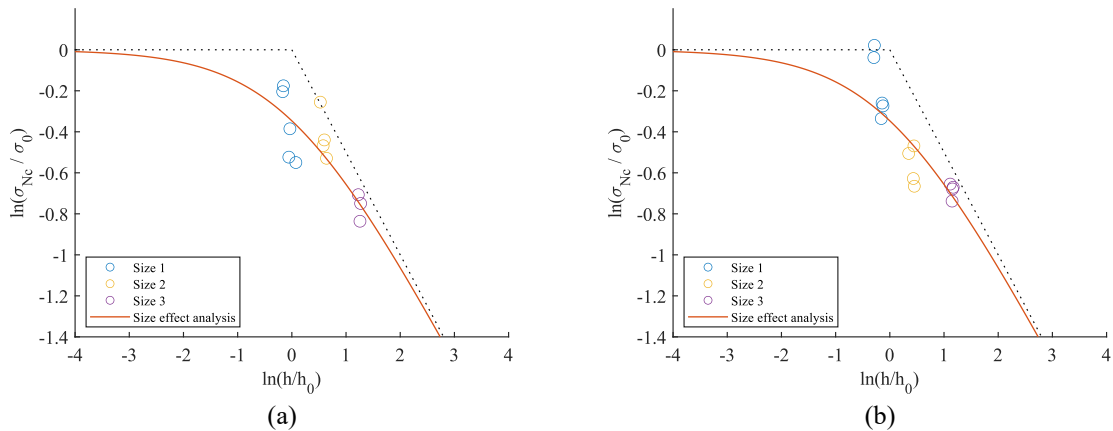

**Figure 8.** Size effect curves. (a) Unstitched specimens. (b) Stitched specimens.

## B. Local fracture behaviors

Local fracture behaviors, as discussed in the methodology section, were analyzed in the form of separation across the midplanes by characterizing the in-plane displacements through the thickness based on the DIC. Details on the analysis method can be found in Ref. [5]. The separation analysis was made for all the specimens; however, only two data sets are presented in this paper. The other data sets will be published in a separate paper. The through-thickness analysis results of unstitched and stitched Size-2 specimens at the peak loads are presented in Figs. 10 and 11, respectively. As discussed in Ref. [5], the in-plane displacement curves near the midplanes are highly affected by fictitious surface tractions indued by DIC speckle paints and thus the plots were extrapolated to obtain separation values which were named as $\Delta u_{\max}$. The separation values were largest at the initial crack tip (i.e., $x$=0) and gradually decreased away from the crack tip along the FPZs for both cases. The unstitched specimen, however, showed larger separation values near the crack tip (see Figs. 9b and 9c) but smaller values away from the tip (Figs. 9d and 9e) compared to the corresponding values of the stitched specimen (see Figs. 10b and 10c and Figs. 10d and 10e, respectively). Additionally, the FPZ sizes of the stitched and unstitched specimens were measured as around 7 mm and 5.5 mm, respectively. This result agrees with the size effect analysis for the effect FPZ size; however, the magnitudes are smaller than $c_f$, which implies partial development of the traction-separation law of the material at the Size-2 dimensions. See Ref. [5] for more discussion on this issue. Lastly, this local analysis also showed that stitching along predicted FPZs could affect the initial quasi-brittle fracture process prior to crack arrestment by increasing the FPZ sizes.

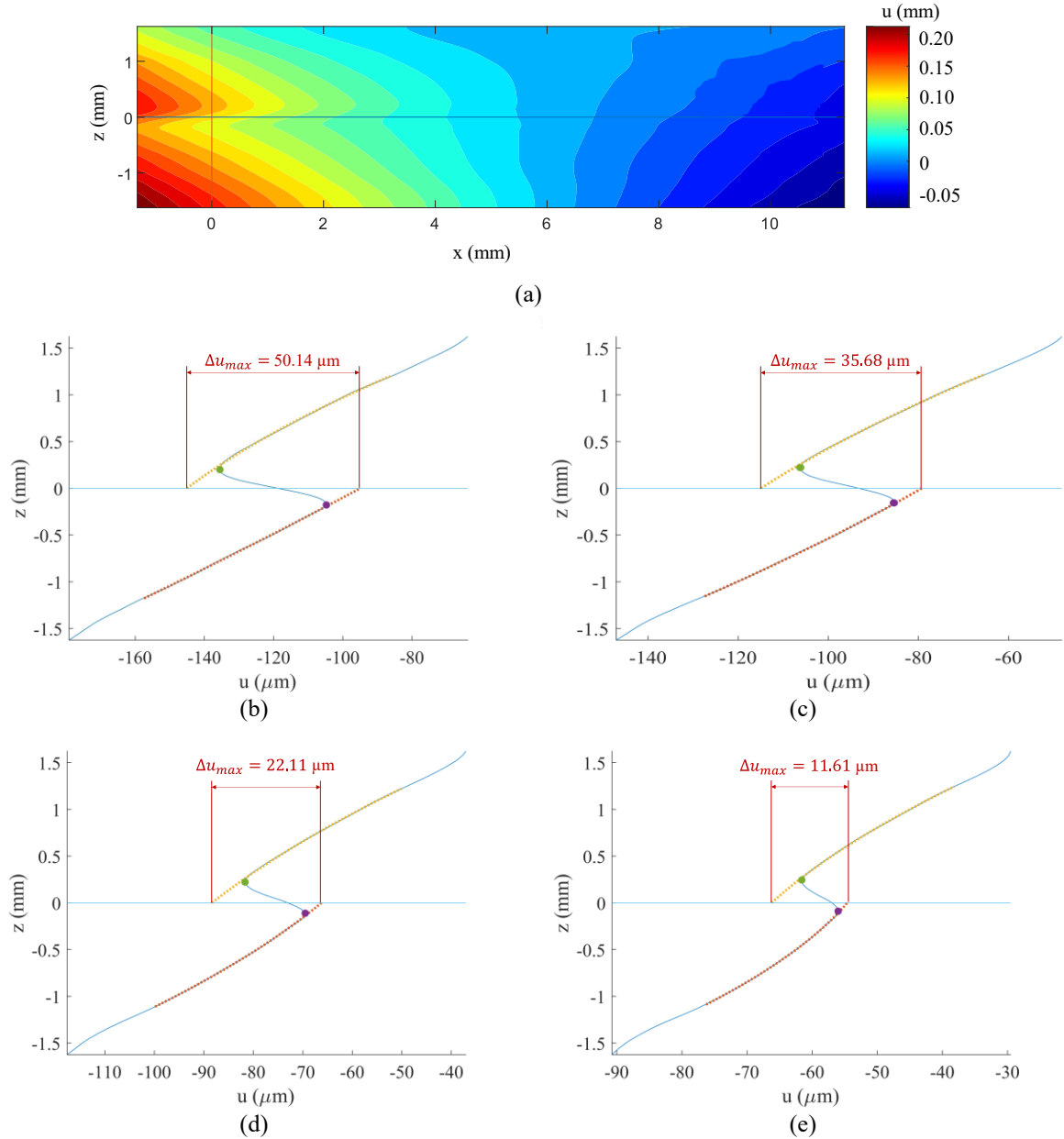

(a)

(b)

(c)

(d)

(e)

**Figure 9. Local fracture analysis for an unstitched Size-2 specimen at peak load. (a) Unstitched specimens. (b) Stitched specimens. (a) Contours of the in-plane displacements $u_1$ on the specimen surface. (b) At $x$=0 (i.e., the initial crack tip). (c) At $x$=1 mm. (d) At $x$=2 mm. (e) At $x$=3 mm.**

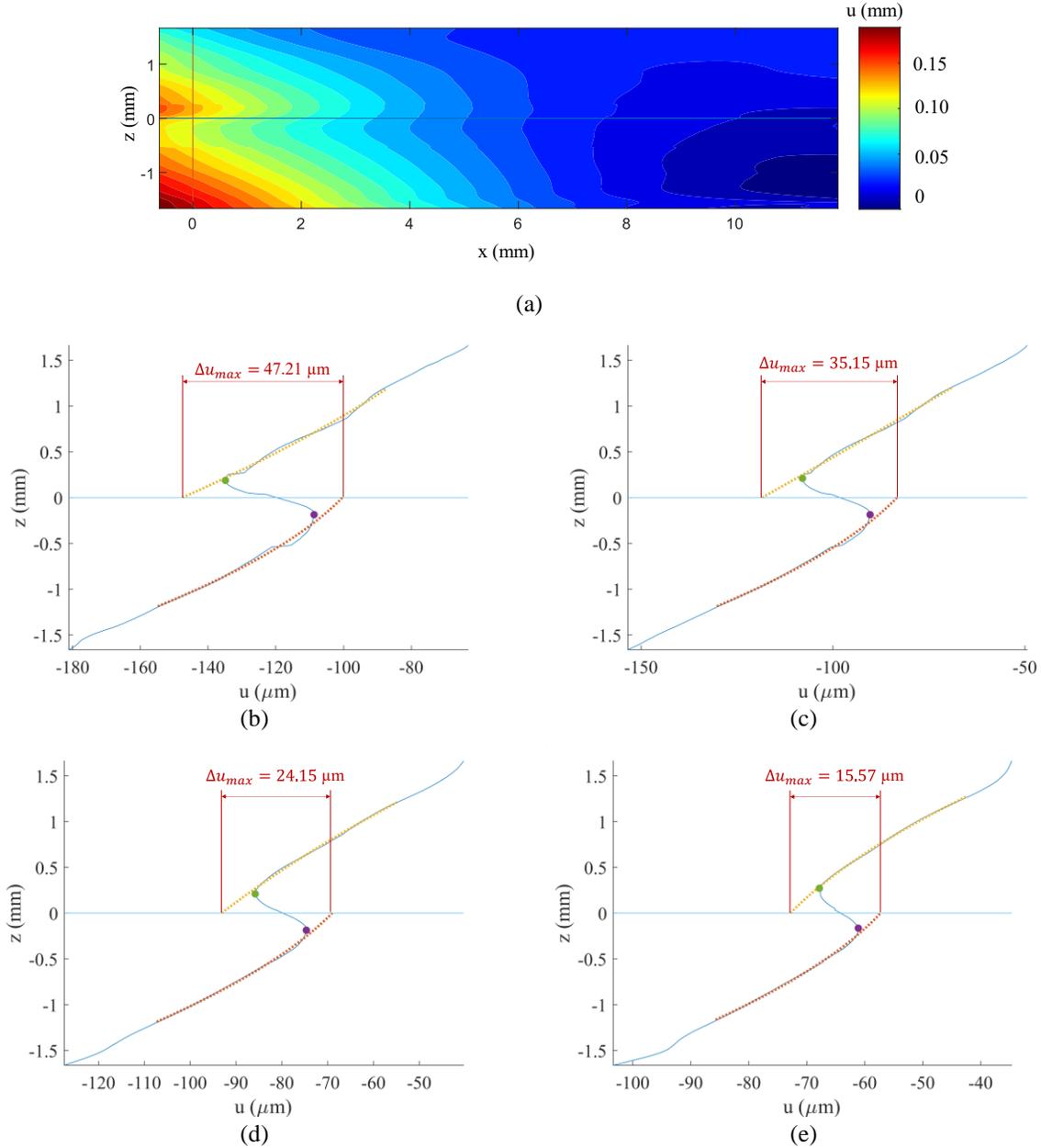

**Figure 10. Local fracture analysis for a stitched Size-2 specimen at peak load. (a) Unstitched specimens. (b) Stitched specimens. (a) Contours of the in-plane displacements $u_1$ on the specimen surface. (b) At $x$=0 (i.e., the initial crack tip). (c) At $x$=1 mm. (d) At $x$=2 mm. (e) At $x$=3 mm.**

## V.  Conclusion

This paper was focused on characterizing the impact of out-of-plane stitches on mode-II interlaminar fracture in geometrically scaled resin-infused composites. The CCM analysis of the experimental data showed larger fracture energy values for the larger size sets. A single fracture energy was obtained through the size effect analysis and was relatively consistent between the stitched and unstitched sets. The effective FPZ length, however, was longer for the stitched case compared to the unstitched set. This phenomenon was observed from the local analysis of the experimental data. Separation across the midplanes was characterized by analyzing in-plane displacements through the thickness based on the DIC data. Similar separation values were observed at the initial crack tips in the stitched and unstitched specimens at the peak loads; however, the FPZ length of the stitched specimen was longer compared to the unstitched one. This could imply that stitching along FPZs could affect initial crack initiation and propagation prior to crack arrestment. The impact of this phenomenon on enhancing interlaminar fracture toughness through out-of-plane stitching will be further investigated in future work. Additionally, the traction-separation law of the material will be experimentally characterized for cohesive zone modeling of damage propagation and crack arrestment in stitched composites. This work will contribute to developing a high-fidelity damage model to predict the structural life of stitched resin-infused composite parts and structures in high-speed aircraft.

## Acknowledgments

The authors gratefully acknowledge the support of the Federal Aviation Administration (FAA) (Award Number: G00003737) and David J. Stanley at FAA.